\documentclass[runningheads]{llncs}
\usepackage{pslatex}
\usepackage{multirow}

\input{psfig.sty}

\begin{document}

\pagestyle{empty}

\mainmatter

\title{Improving Term Frequency Normalization for Multi-topical Documents, and Application to Language Modeling Approaches}

\titlerunning{Lecture Notes in Computer Science}

\author{Seung-Hoon Na\inst{1} \and In-Su Kang\inst{2} \and Jong-Hyeok Lee\inst{1}}


\authorrunning{Improving Term Frequency Normalization for Multi-topical Documents}

\institute{POSTECH,Pohang,South Korea,\{nsh1979,jhlee\}@postech.ac.kr \\
\and KISTI,Daejeon,South Korea,dbaisk@kisti.re.kr}


\maketitle

\begin{abstract}
Term frequency normalization is a serious issue since lengths of documents are various. Generally, documents become long due to two different reasons - verbosity and multi-topicality. First, verbosity means that the same topic is repeatedly mentioned by terms related to the topic, so that term frequency is more increased than the well-summarized one. Second, multi-topicality indicates that a document has a broad discussion of multi-topics, rather than single topic. Although these document characteristics should be differently handled, all previous methods of term frequency normalization have ignored these differences and have used a simplified length-driven approach which decreases the term frequency by only the length of a document, causing an unreasonable penalization. To attack this problem, we propose a novel TF normalization method which is a type of partially-axiomatic approach. We first formulate two formal constraints that the retrieval model should satisfy for documents having verbose and multi-topicality characteristic, respectively. Then, we modify language modeling approaches to better satisfy these two constraints, and derive novel smoothing methods. Experimental results show that the proposed method increases significantly the precision for keyword queries, and substantially improves MAP (Mean Average Precision) for verbose queries.
\end{abstract}

\section{Introduction}
The highly-performed retrieval models rely on two different factors - TF (term frequency) and IDF (inverse document frequency). Among them, TF factor becomes a non-trivial, since long-length documents may increase term frequency, different to short-length ones, so that the naive estimation of term frequency would not be successful. Thus, term frequency of long-length documents should be seriously considered. Regarding this, Singhal observed the following two different types of reasons for making the length of a document long \cite{singhal96} \footnote{
Robertson and Walker mentioned two types of reasons as scope hypothesis and verbosity hypothesis, respectively \cite{robertson94}.}.

\begin{enumerate}
  \item High term frequency: The same term repeatedly occurs in a long-length document. As a result, the term frequency factors may be large for long documents, increasing the average contribution of its terms towards the query-document similarity.
  \item More terms: Long-length document has large size of vocabulary. This increases the number of matches between a query and a long document, increasing the query-document similarity, and the chances of retrieval of long documents in preference over shorter documents.
\end{enumerate}
Without loss of meaning, we can conceptualize these two reasons as verbosity and multi-topicality. First, verbosity means that the same topic is repeatedly mentioned by terms related to the topic, making term frequencies high. Second, multi-topicality indicates that a document has a broad discussion of multi-topics, rather than single topic, making more terms. Using these concepts, we divide long-length documents into two different ideal types - verbose documents and multi-topical documents. Verbose document is the document which becomes long mainly due to verbosity, rather than multi-topicality, while multi-topical document is the document which follows typical characteristics of multi-topicality, rather than verbosity.

Singhal pre-assumed that long-length documents should be penalized regardless of whether or not their types are verbosity (or multi-topicality) \cite{singhal96}. Basically, their approach belongs to a simplified length-driven method which decreases the term frequency of all long-length documents according to documents' length factor only. However, we insist that this Singhal's pre-assumption would be failed. We argue that the penalization should be applied to verbose document only, not to multi-topical document. As a main reason, terms in a multi-topical document are less repeated than ones in a verbose document, since the length of the multi-topical document is increased due to its broad topics. However, Singhal missed this point that these types of documents should be differently handled. Therefore, the retrieval function adopting Singhal's penalization will make multi-topical documents unreasonably less-preferred, causing an unfair retrieval ranking.

To clearly support our argument for verbose document and multi-topical document, we will exemplify two different situations to discuss different tendencies of term frequencies in verbose document and multi-topical document. First, let us examine the situation by considering two different document samples of $D_1$ and $D_2$ which have the same term frequency ratio.
\begin{center}
\begin{tabular}{|p{150pt}|}
\hline
$D_1$: Language modeling approach
\\\hline
\end{tabular}
\begin{tabular}{|p{150pt}|}
\hline
$D_2$: Language modeling approach Language modeling approach
\\\hline
\end{tabular}
\end{center}

$D_2$ is twice the concatenation of $D_1$. Suppose that a query is given by ``language modeling approach''. Then, a question arises as ``which one of $D_1$ and $D_2$ is more relevant?''. By comparing the contained information, we know that two documents have the exactly same contents, although the length of $D_2$ is twice than that of $D_1$. Thus, $D_1$ and $D_2$ should have the same relevance score. However, the absolute term frequency of $D_2$ is twice than that of $D_1$, thus, the naive TF $\cdot$ IDF prefers $D_2$ to $D_1$. To avoid this unfair comparison, we should introduce a TF normalization. To this end, suppose that $l$ is the length of documents, and $tf$ is the term frequency of a query term. Then, one reasonable strategy of TF normalization is to use $tfn = tf/l$, instead of $tf$. Then, the modified TF $\cdot$ IDF produces the same score for $D_1$ and $D_2$. Note that Singhal's pivoted length normalization will also well-work since $tfn$ can be well-reflected in Singhal's original formula. Remark that $D_2$ is a verbose document, not a multi-topical document, which is the main reason for the success of the normalization. Now, we examine the second situation by considering a multi-topical document sample $D_3$, which contains all topics of $D_1$ and $D_2$ as a subpart.
\begin{center}
\begin{tabular}{|p{150pt}|}
\hline
$D_3$: Information retrieval model \\
Language modeling approach
\\\hline
\end{tabular}
\end{center}
Here, $D_3$ describes a broad topic - ``information retrieval model'', and contain ``language modeling approach'' as a subtopic. Again, suppose that the same query of ``language modeling approach'' is given. Consider the question about ``what relevance score should assigned to $D_3$ be, compared with $D_1$ and $D_2$?''. $D_3$ contains all contents of $D_1$ and $D_2$, although $D_3$ is different from $D_1$ and $D_2$. In this case, if user sees $D_3$, he or she would think that $D_3$ is also relevant, because all relevant content - $D_1$ - is embedded to $D_3$. From this viewpoint, $D_3$ should have the same score as $D_1$ and $D_2$ (due to a partial relevance). However, if we apply the previous version of TF-normalization (i.e. $tfn = tf/l$) to $D_3$, then $D_3$ is much-less preferred to $D_1$ and $D_2$, since its term frequency of a query term is the same as $D_1$ but its length is twice than that of $D_1$. Of course, Singhal's method will assign less-score to $D_3$ than $D_1$ and $D_2$. The mean reason of this failure is that $D_3$ is not a verbose document but a multi-topical document. This result means that TF normalization problem is more complex, at least requiring the different strategies according to types of long-length documents. To avoid the unreasonable penalization for multi-topical ones, TF normalization problem should be more deeply re-investigated by discriminating multi-topical documents from verbose documents.

To obtain a more accurate TF normalization, we propose a novel TF normalization method which is a type of axiomatic approach. We try to modify language modeling approach as a case study without the loss of its elegance and principle. To this end, we first formulate two constraints that the retrieval scoring functions should satisfy for verbose and multi-topical documents, respectively. Then, we present the analysis result that previous language modeling approaches do not sufficiently satisfy these constraints. After that, we modify the language modeling approaches such that better satisfy these two constraints, derive a novel smoothing methods, and evaluate the proposed ones.

\section{Formal Constraints of New TF Normalization, and Analysis of Previous Language Modeling Approaches}
\subsection{Constraints}
From now on, we assume that $\tau(D)$ is a measurement for calculating the number of topics in document $D$. We define \emph{K-verbosity} and \emph{N-topicality} as follows.

\textbf{Definition (K-verbosity)}: Suppose that $D_1$ and $D_2$ are given. Let $tf_1(w)$ and $tf_2(w)$ be the term frequency of term $w$ in $D_1$ and $D_2$, respectively. For all term $w$, if $tf_2(w)$ = $K \cdot tf_1(w)$ and $\tau(D_1)$ = $\tau(D_2)$, then $D_2$ has K-verbosity to $D_1$ or $D_2$ is \emph{K-verbose} to $D_1$.

\textbf{Definition (N-topicality)}: Suppose that $D_1$ and $D_2$ are given as $\tau(D_2)$ = $N \cdot \tau(D_1)$. Let $l_1$ and $l_2$ be the length of $D_1$ and $D_2$, respectively. If for all term $w$ in $D_1$, $tf_2(w)/l_2$ = $tf_1(w)/l_1/N$, then $D_2$ has N-topicality to $D_1$ and $D_2$ is \emph{N-topical} to $D_1$.

In our three samples from the introduction, $D_2$ has 2-verbosity to $D_1$, and $D_3$ has 2-topicality to $D_1$. Remind that we have argued that $D_1$, $D_2$ and $D_3$ should have the same relevance score. This argument can be re-formulated to following two constraints - VNC and TNC which the retrieval function should satisfy for two cases when one document has K-verbosity and N-topicality to another document, respectively. Let $score(Q,D)$ be a similarity function between a document $D$ and a query $Q$.

\textbf{VNC (Verbosity Normalization Constraint)}: Suppose a pair of $D_1$ and $D_2$. If $D_2$ is K-verbose to $D_1$, then $score(Q,D_1)$ = $score(Q,D_2)$.

\textbf{TNC (Topicality Normalization Constraint)}: Suppose a pair of $D_1$ and $D_2$.  If $D_2$ is N-topicality to $D_1$, then $score(Q,D_1)$ = $score(Q,D_2)$.

These constraints can be directly utilized to derive a new class of retrieval function as Fang's exploration \cite{fang04}. Originally, Fang formulated two constraints related to term frequency - LNC1 and LNC2 \cite{fang04}. Among them, LNC2 is highly relevant to VNC, where VNC is a more specific constraint - VNC entails LNC2, not vice versa. TNC is a new constraint which is not connected to Fang's any constraint. Note that our exploration of a retrieval function is different from Fang's one. We focus on only few constraints related to our issue, without identifying all constraints. Then, we select as the backbone model one among a previous well-performed retrieval model, and modify it to better satisfy the focused few constraints, without losing the elegance and the principle of the original model. In this regard, our exploration method belongs to the \emph{partially-axiomatic} approach - 1) using partial constraints rather than full constraints, 2) using the restricted functional space which the backbone retrieval model can allows, rather than relying on full functional space. In contrast, Fang's approach is the \emph{fully-axiomatic} approach \cite{fang04,fang05}. In Fang's approach, full constraints are completely identified as well as the focused constraints. A new class of retrieval function is explored as one in separate functional space which is not related to previous retrieval models. However, the fully-axiomatic approach such as Fang's exploration approach requires un-principled heuristics which are not derived from a well-designed retrieval model. A partially-axiomatic approach doesn't need to discard the well-founded retrieval model such as language modeling approach, enabling us to pursue a more elaborated retrieval model, without losing its mathematical elegance and principles.

\subsection{Analysis of Language Modeling Approaches}
We selected the language modeling approaches as the backbone retrieval model \cite{ponte}. Our goal is to modify the language modeling approaches such that better satisfies the proposed two constraints - VNC and TNC.  We investigate two popular smoothing methods - Jelinek-Mercer smoothing (JM) and Dirichlet-prior smoothing (Dir) \cite{zhai01}. Before modifying them, we begin by discussing whether or not each smoothing method satisfies VNC and TNC in this subsection. Notations used in this paper are summarized as follows:

\begin{tabular}{|ll|}
\hline
$Q$ & A given query  \\
$tf_D(w)$ & Term frequency of $w$ in document $D$ \\
$l_D$ & Length of document $D$ \\
$tf_C(w)$ & Term frequency of $w$ of collection \\
$l_C$ & Total term frequency of collection \\
$\theta_D$ & Smoothed document language model of $D$ \\
$\hat{\theta}_D$ & Unsmoothed document language model of $D$ (MLE) \\
$\theta_C$ & Collection language model (MLE) \\\hline
\end{tabular}

\subsubsection{Analysis of Jelinek-Mercer Smoothing}
In JM (Jeliner-Mercer Smoothing), a smoothed document model is obtained by the interpolation of MLE (Maximum Likelihood Estimation) of a document model and the collection model as follows \cite{zhai01}:
\begin{equation}
P(w|\theta_D) = (1-\lambda) P(w|\hat{\theta}_D) + \lambda P(w|\theta_C)
\label{eq_JM_smoothing}
\end{equation}
where $\lambda$ is a smoothing parameter. By using JM, $score(Q,D)$, the similarity score of document $D$ for query $Q$ can be written by using only query-matching terms as follows:
\begin{equation}
score(Q,D) = \sum_{w \in Q} \mbox{log} \left( \frac{1-\lambda}{\lambda} \frac{P(w|\hat{\theta}_D)}{P(w|\theta_C)} + 1 \right)
= \sum_{w \in Q} log \left( \frac{1-\lambda}{\lambda} \frac{tf_D(w)}{l_D}\frac{l_C}{tf_C(w)} + 1 \right)
\label{eq_JM_similarity_function}
\end{equation}
Our analysis of whether or not JM satisfies VNC and TNC is given as follows:
\begin{enumerate}
  \item JM satisfies VNC: Suppose that $D_2$ is K-verbose to $D_1$. Then, MLEs of two document models are the same, resulting in the same scores.
  \item JM does not satisfy TNC: Generally, JM prefers normal documents to multi-topical documents, regardless of our definition of topicality measurement $\tau$. This proof is skipped.
\end{enumerate}

\subsubsection{Analysis of Dirichlet-Prior Smoothing}
In Dir (Dirichlet-prior smoothing), a smoothed document model is estimated as posterior model when taking $\mu P(w|\theta_C)$ as a prior probability of term $w$ as follows \cite{zhai01}:
\begin{equation}
P(w|\theta_D) = \frac{tf_D(w) + \mu P(w|\theta_C)}{l_D + \mu}
\label{eq_Dirichlet_smoothing}
\end{equation}
The equation is rewritten by
\begin{equation}
P(w|\theta_D) = \frac{l_D}{l_D + \mu} P(w|\hat{\theta}_D) + \frac{\mu}{l_D + \mu} P(w|\theta_C)
\label{eq_Dirichlet_smoothing_mixture_version}
\end{equation}
If we set $\lambda_D$ by $\mu/(l_D + \mu)$, then Dir is equivalent to JM-style smoothing using document-specific smoothing parameter $\lambda_D$. $score(D,Q)$ based on Dir is formulated as follows:
\begin{displaymath}
score(D,Q) = \sum_{w \in Q} \mbox{log} \left( (1-\lambda_D) \frac{P(w|{\hat\theta}_D)}{P(w|\theta_C)} + \lambda_D \right)
\end{displaymath}
The analysis on whether or not Dir satisfies VNC and TNC is somewhat complicated, due to its document-specific smoothing parameter. We can easily show that Dir does not satisfy VNC and TNC. The following lists up the analysis result. \begin{enumerate}
  \item Dir doesn't satisfy VNC: Generally, Dir makes inconsistent preferences according to whether or not a query term is topical. For a topical query term, Dir assigns the more score for verbose documents than normal documents. For a non-topical query terms, Dir assigns the less score for verbose documents than normal documents. The detailed proof is skipped.
  \item Dir doesn't satisfy TNC: The detailed proof is skipped.
\end{enumerate}

\section{Modification of Previous Retrieval Models}
In the previous section, we have shown that two different smoothing methods do not satisfy two constraints well. In this section, we introduce the measurement of the number of topics, and modify the previous retrieval model such that it better satisfies VNC and TNC.

\subsection{Measurement of The Number of Topics}
To figure out which measurement $\tau(D)$ is acceptable to calculate the number of topics in document $D$, we propose two simple measurements for $\tau(D)$ - The first one is \emph{vocabulary size}, and the second one is \emph{information quantity}.

\textbf{Vocabulary Size}: Generally, as there are more terms, a given document has more topics. Based on this idea, we can use the vocabulary size - $\nu(D)$ - which indicates the number of unique terms in a given document, as a measurement for the number of topics.

\textbf{Information Quantity}: Even though the vocabulary size is simple and reasonable, it cannot discriminate the mainly topical terms from the causally-occurred terms. When using the vocabulary size, the number of topics may be unreasonably increased due to causally occurred terms. As for an alternative measurement, we consider the entropy-driven value. Remind that entropy means the uncertainty of a generated sample. Entropy has the following positive properties for resolving the limitation of the vocabulary size. 1) As the number of possible events increases, entropy becomes larger. Here, events correspond to terms, hence the more terms are, the larger the entropy is likely to be. Thus, when a document has more topics, the content of the document can be described in more various ways, resulting in a larger entropy value. 2) Term generative probability of a document is used as the weight for calculating entropy value. As a term has more large probability, it makes more contribution to the final-entropy value. This property allows us to differentiate the effects of mainly topical terms and causally occurred terms.

The information quantity - $\varepsilon(D)$ - is defined as an exponential function of entropy of a document as follows:
\begin{displaymath}
\tau(D) = \varepsilon(D) = \mbox{exp}\left( -\sum_w P(w|\theta_D) \mbox{log} P(w|\theta_D) \right)
\end{displaymath}

\textbf{Some Useful Definitions}: We define some useful notations. Let us define the normalized measurement of the number of topics - $\tau'(D)$ -, and define the \emph{informative verbosity} - $\omega(D)$ - as follows:

\begin{displaymath}
\tau'(D) = \tau(D)/{\tilde{\tau}},\qquad \omega(D) = l_D/\tau(D)
\end{displaymath}
where $\tilde{\tau}$ is the mean of $\tau(D)$ for all documents in a given test collection. Note that the informative verbosity indicates the average term frequency per unit information.

\subsection{Modification of JM}
\subsubsection{First Modification of JM}
Since JM exactly satisfies VNC, we would try to modify JM to additionally support TNC. The core idea of the modification of JM smoothing is a pseudo document. The pseudo document mainly consists of relevant parts to a query, which is constructed by extracting relevant parts from non-relevant parts. Then, the score of a document is calculated by using the pseudo document model, instead of original document model.

Thus, the pseudo-document makes us take a dynamic viewpoint of document representation where a document is dynamically changed according to a query. Note that a pseudo document is an imaginary concept, which is not really constructed at real time. All we require is generative probabilities for query terms from the pseudo document model.

To estimate probability of query terms in a pseudo document, we simplify the estimation problem by using probability in original document. In other words, for terms in the pseudo document having non-zero probabilities, their probabilities are assumed to be proportional to the probabilities of terms in the original document. As a result, the estimation problem is completed only if we determine the length of the pseudo document from the original length $l_D$.

Intuitively, the length of the pseudo document will be smaller, as topics are more. This intuition makes the length of the pseudo document proportional to $l_D/\tau(D)$. Thus, if $\theta_{{Pseudo}(D)}$ is the language model of pseudo document, then the probability of pseudo document model is
\begin{displaymath}
P(w|\theta_{Pseudo(D)}) \propto tf_D(w) / l_D / \tau(D) = tf_D(w) \cdot \tau(D) / l_D
\end{displaymath}
It is rewritten by using $\tau'(D)$ instead of $\tau(D)$, and the constant $K$ as follows:
\begin{displaymath}
P(w|\theta_{Pseudo(D)}) = K \cdot tf_D(w) \cdot \tau'(D) / l_D
\end{displaymath}
If we assume that the constant $K$ is independent to any document and query, then $K$ is not a tuning parameter since it can be included in smoothing parameter $\lambda$.

Let us derive a modified JM by substituting the original document model to this pseudo document model in Eq. (\ref{eq_JM_similarity_function}). Then, $score(Q,D)$ is reformulated as follows:
\begin{equation}
score(Q,D) = \sum_{w \in Q} \mbox{log} \left( \frac{1-\lambda_0}{\lambda_0} \frac{K \cdot \tau'(D) \cdot tf_D(w)}{l_D} \frac{l_C}{tf_C(w)} + 1 \right)
\label{eq_JMV_similarity_function}
\end{equation}
where $\lambda_0$ is another smoothing parameter for the pseudo document model. Since $K$ is independent to any document and query, we can select $\lambda$ such that  $(1-\lambda_0) K : \lambda_0$ is $(1-\lambda): \lambda$, in order to eliminate constant $K$. Then, Eq. (\ref{eq_JMV_similarity_function}) is re-written by
\begin{equation}
score(Q,D) = \sum_{w \in Q} \mbox{log} \left( \frac{1-\lambda}{\lambda} \frac{\tau'(D) \cdot tf_D(w)}{l_D} \frac{l_C}{tf_C(w)} + 1\right)
\label{eq_JMV_similarity_function_elim_K}
\end{equation}
By using MLE of the original document model $P(w|{\hat\theta}_D)$, Eq. (\ref{eq_JMV_similarity_function_elim_K}) is rewritten by
\begin{equation}
score(Q,D) = \sum_{w \in Q} \log \left( \frac{1-\lambda}{\lambda} \tau'(D) P(w|{\hat\theta}_D) \frac{l_C}{tf_C(w)} + 1\right)
\label{eq_JMV_similarity_function_final}
\end{equation}
Eq. (\ref{eq_JMV_similarity_function_final}) is the final modified JM, which is called JMV. JMV satisfies both of VNC and TNC.
\begin{enumerate}
  \item JMV satisfies VNC: Let $D_2$ be K-verbose to $D_1$. Then, $\tau(D_1)$ = $\tau(D_2)$ and $P(w|D_1)$ = $P(w|D_2)$. Thus, $score(Q,D_1)$ = $score(Q,D_2)$.
  \item JMV satisfies TNC: Let $D_3$ be N-topical to $D_1$. Then, $\tau(D_3)$ = $ N \tau(D_2)$ and $P(w|D_1)$ = $N P(w|D_3)$. It makes that $\tau(D_3) P(w|D_3)$ = $\tau(D_1) P(w|D_1)$. Therefore, $score(Q,D_1)$ = $score(Q,D_3)$.
\end{enumerate}

\subsubsection{Second Modification of JM}
In our preliminary experiments, we found that JMV performs well for keyword queries (i.e. title query), but is not reliable for verbose queries (i.e. description query), by showing serious sensitivity according to smoothing parameter $\lambda$. To discuss the reason of this result, we focus on the main differences of keyword query and verbose query. First, there are common terms in a verbose query. Different from topical terms, common terms can be shared by all topics. A common term always verbosely acts regardless of verbose documents and multi-topical documents. Thus, the previous TF normalization would prefer multi-topical documents for queries including common terms. Second, verbose queries often contain noise terms such as ``relevant'', ``find'' and ``documents''. When a document has more topics, it will increase the chance of existence of such noise terms. However, when our previous TF normalization is applied, noise term becomes very serious, because the number of topics is further multiplied to the normalized term frequency. Thus, the previous TF normalization would increase the scores of multi-topical documents for noise queries. These two differences may be the reason why Singhal et. al. penalized even multi-topical documents, as well as verbose documents \cite{singhal96}. However, we already discussed that their approach is not acceptable to topical terms.

To handle the problems of verbose-type queries, our TF normalization should be restricted to only document-specific terms, not to noise terms or common terms. As a query term is more topical term in a given document, we hope to perform more TF normalization, and vice versa. To this end, we define $s(w,D)$ as term specificity of $w$ in document $D$. As for $s(w,D)$ this paper uses a probabilistic metric $P(D|w)$ which is defined as follows:
\begin{displaymath}
s(w,D) = P(D|w) = \frac{\lambda_s P(w|{\hat{\theta}}_D)}{\lambda_s P(w|{\hat{\theta}}_D) + (1-\lambda_s) P(w|\theta_C) }
\end{displaymath}
where $\lambda_s$ is an additional smoothing parameter, which has 0.25 as the default value.  By using the term specificity $s(w,D)$, we newly modify the pseudo document model as follows:
\begin{equation}
P(w|\theta_{Pseudo(D)}) = K \cdot tf_D(w) \cdot \tau'(D)^{P(D|w)}/l_D
\label{eq_JMV2_pseudo_document_model}
\end{equation}
Since $P(D|w)$ is between 0 and 1, the normalization is perfectly reflected when $P(D|w)$ is 1, while it is weaken as $P(D|w)$ is close to 0. One problem arises when $\tau'(D)$ is smaller than 1. In this case, as $P(D|w)$ is larger, the effect of normalization becomes weaker. To resolve this problem, we considered the exceptional TF normalization, making the normalization proportional to $P(D|w)$ even when $\tau'(D)$ is smaller than 1. In preliminary experiments, we found that the final retrieval performance is almost not changed, even after the exceptional TF normalization is applied. Thus, we select Eq. (\ref{eq_JMV2_pseudo_document_model}) for second modification. We call it JMV2.

\section{Modification of Dir}
Our goal for Dir modification is to provide VNC. We introduce the concept of pseudo document model to modify Dir. Different from the pseudo document for JM modification that consists of query-relevant parts only, the pseudo document for Dir modification consists of all topics in the original document, but has a different length from the original length. Note that the change of the length only makes different models, since the smoothed model - $P(w|\theta_D)$ - is different according to the document length. In fact, the length-dependence was the main reason why Dir does not satisfy VNC.

We assume that the pseudo document model is proportional to original MLE document model. In addition, we set the length of the pseudo document by $\tau(D)$. Remind that informative verbosity - $\omega(D)$ - is defined as $l_D/\tau(D)$. That is, the pseudo document with length of $\tau(D)$ compacts the original document with length $l_D$ by $\omega(D)$ time. Therefore, each term $w$ of document $D$ has the following term frequency in the pseudo document.
\begin{equation}
tf_{Pseudo(D)}(w) = tf_D(w) / \omega(D)
\label{eq_DirV_pseudo_tf}
\end{equation}

As a result, the pseudo document model becomes length-independent model, even though MLE of pseudo document model is the same as the original document model. By using pseudo document model, Dir produces the following smoothed model.
\begin{equation}
P(w|\theta_{Pseudo(D)}) = \frac{tf_{Pseudo(D)}(w) + \mu P(w|\theta_C)}{\tau(D) + \mu}
\label{eq_DirV_pseudo_document_model}
\end{equation}
By substituting Eq. (\ref{eq_DirV_pseudo_tf}) to Eq. (\ref{eq_DirV_pseudo_document_model}), Eq. (\ref{eq_DirV_pseudo_document_model}) becomes
\begin{equation}
P(w|\theta_{Pseudo(D)}) = \frac{\tau(D)}{\tau(D) + \mu}P(w|{\hat\theta}_D) + \frac{\mu}{\tau(D) + \mu}P(w|\theta_C)
\end{equation}
This final modified model can be viewed as JM-style smoothing using document-specific smoothing paramter $\lambda_D$ with $\mu/(\tau(D)+\mu)$, which is not dependent to the length any more. We call this modification DirV. We can easily prove that DirV additionally satisfies VNC.
\begin{enumerate}
  \item DirV satisfies VNC: Let $D_2$ be K-verbose to $D_1$. Then, two MLE models are equal (i.e $P(w|\theta_{D_1})$ = $P(w|\theta_{D_2})$). $\lambda_{D_1}$ is $\lambda_{D_2}$ since $\tau(D_1)$ and $\tau(D_2)$ are the same. Thus, DirV gives the same score for $D_1$ and $D_2$.
  \item DirV does not satisfy TNC: For DirV, we do not have a special consideration for supporting TNC.
\end{enumerate}

\section{Experimentation}

\subsection{Experimental Setting}
For evaluation, we used five TREC test collections. The standard method was applied to extract index terms; We first separated words based on space character, eliminated stopwords, and then applied Porter's stemming. Table \ref{tbl_collection_summary} summarizes the basic information of each test collection. In columns, \#\emph{Q}, \emph{Topics}, \#\emph{R}, \#\emph{Doc}, \emph{avglen}, and \#\emph{Terms} are the number of topics, corresponding query topic IDs, the number of relevant documents, the number of documents, the average length of documents, and the number of terms, respectively.

\begin{table}
\centering \caption{Collection summaries}
\begin{tabular*}{0.9\textwidth}%
{@{\extracolsep{\fill}}|c|c|c|c|c|c|c|}
\hline
\emph{Collection} & \# \emph{Q}	& \emph{Topics}	& \# \emph{R} & \# \emph{Doc}	& \emph{avglen}	 & \# \emph{Term}	 \\\hline
TREC7	& 	 50	& 350-400	& 4,674 	&	\multirow{2}{*}{528,155}	& \multirow{2}{*}{154.6}	& \multirow{2}{*}{970,977}				\\\cline{1-4}
TREC8	&		 50	& 401-450	& 4,728 	&		&         &         																															 \\\hline
WT2G	& 	 50	& 401-450	& 2,279 	&	247,491	& 254.99	& 2,585,383																													 \\\hline
TREC9	& 	 50	& 451-500	 & 2,617  &	\multirow{2}{*}{1,692,096}	& \multirow{2}{*}{165.16}	& \multirow{2}{*}{13,018,003}		\\\cline{1-4}
TREC10	&  50	& 501-550	& 3,363 	&		     &	      &                																								  \\\hline
\end{tabular*}
\label{tbl_collection_summary}
\end{table}
According to Zhai's work \cite{zhai01}, we used the following three different types of queries:

1) \textbf{Short keyword (SK)}: Using only the title of the topic description.

2) \textbf{Short Verbose (SV)}: Using only the description field (usually one sentence).

3) \textbf{Long Verbose (LV)}: Using the title, description and the narrative field (more than 50 words on average).

As for retrieval evaluation, we used MAP (Mean Average Precision), Pr@5 (Precision at 5 documents), and Pr@10 (Precision at 10 documents).


\subsection{Experimental Results}
Table \ref{tbl_experimental_results} shows the best performances (MAP, Pr@5, Pr@10) of DirV and JMV2, compared with Dir. As for topic measurement $\tau(D)$, we selected the information quantity ($\varepsilon(D)$) since JMV2 and DirV using the information quantity is better than those using vocabulary size. We used MLE (Maximum Likelihood Estimation) for $P(w|\theta_D)$ to calculate the information quantity without any smoothing. We selected Dir as the baseline due to its superiority over JM in all test collections. To obtain the best performance of each run, we searched 20 different values between 0.01 and 0.99 for $\lambda$, and 22 values between 100 and 30,000 for $\mu$. To check whether or not the proposed method (DirV and JMV2) significantly improves the baseline, we performed the Wilcoxon sign ranked test to examine at 95\% and 99\% confidence levels. We attached $\dagger$ and $\ddagger$ to the performance number of each cell in the table when the test passes at 95\% and 99\% confidence level, respectively. The results are summarized as follows:
\begin{enumerate}
\item DirV significantly improves MAP of Dir for verbose type of query (SV and LV). Exceptionally, TREC10 did not show an improvement for verbose type of query.
\item DirV does not significantly improve MAP of Dir for keyword type of query (SK), but improves precisions (Pr@5 or Pr@10). Especially, on TREC7 and TREC8, Pr@10 is significantly improved over Dir. Although other test collections do not statistically show a significant improvement, there is large portion of the numerical increase.
\item DirV or JMV2 show improvement on a specific test collection even for keyword type of query. For DirV, TREC10 is such a collection by showing a significant improvement of MAP. For JMV2, WT2G is such a test collection.
\item Overall, DirV is slightly better than JMV2 in most of test collections. WT2G is an exceptional collection to show that JMV2 significantly improves DirV.
\end{enumerate}

\begin{table}[t]
\centering \caption{Performances of Dir, DirV and JMV2 (MAP, Pr@5, Pr@10). Bold faced numbers indicate runs showing significant improvement over Dir. }
\begin{tabular*}{1.0\textwidth}%
{@{\extracolsep{\fill}}|c|c|c|c|c|c|c|c|c|c|}\hline
\multirow{2}{*}{\textbf{MAP}} & \multicolumn{3}{c|}{Dir} &    \multicolumn{3}{c|}{DirV} & \multicolumn{3}{c|}{JMV2} \\\cline{2-10}
 & SK & SV & LV & SK & SV & LV & SK & SV & LV \\\hline
TREC7  & 0.1786 & 0.1790 & 0.2209 & 0.1835 & \textbf{0.1967}$\ddagger$ & \textbf{0.2348}$\ddagger$ & 0.1825 & \textbf{0.1926}$\dagger$ & 0.2250 \\\hline
TREC8  & 0.2481 & 0.2294 & 0.2598 & 0.2492 & \textbf{0.2393}$\ddagger$ & \textbf{0.2621}$\ddagger$ & \textbf{0.2505}$\dagger$ & \textbf{0.2354}$\dagger$ & 0.2500 \\\hline
WT2G  & 0.3101 & 0.2854 & 0.2863 & 0.3125 & \textbf{0.3103}$\ddagger$ & \textbf{0.3267}$\ddagger$ & \textbf{0.3278}$\ddagger$ & \textbf{0.3112}$\ddagger$ & \textbf{0.3263}$\ddagger$ \\\hline
TREC9 & 0.2038 & 0.1990 & 0.2468 & 0.2040 & \textbf{0.2336}$\ddagger$ & \textbf{0.2581}$\ddagger$ & \textbf{0.2068} & \textbf{0.2245}$\ddagger$ & 0.2494 \\\hline
TREC10 & 0.1950 & 0.1865 & 0.2347 & \textbf{0.2049}$\dagger$ & 0.2248 & 0.2640 & 0.2091 & \textbf{0.2133}$\dagger$ & 0.2555 \\\hline\hline
\multirow{2}{*}{\textbf{Pr@5}} & \multicolumn{3}{c|}{Dir} &    \multicolumn{3}{c|}{DirV} & \multicolumn{3}{c|}{JMV2} \\\cline{2-10}
 & SK & SV & LV & SK & SV & LV & SK & SV & LV \\\hline
TREC7  & 0.4400 & 0.4280 & 0.5240 & 0.4560 & \textbf{0.4840}$\dagger$ & \textbf{0.5680}$\dagger$ & 0.4680 & \textbf{0.4920}$\dagger$ & \textbf{0.5800}$\dagger$ \\\hline
TREC8  & 0.4920 & 0.4320 & 0.5120 & 0.5120 & \textbf{0.5040}$\dagger$ & 0.5360 & \textbf{0.5240}$\ddagger$ & 0.4880 & 0.5280 \\\hline
WT2G  & 0.5160 & 0.5120 & 0.5280 & 0.5360 & 0.5520 & \textbf{0.5720}$\dagger$ & 0.5400 & 0.5560 & \textbf{0.5920}$\dagger$ \\\hline
TREC9 & 0.3000 & 0.3480 & 0.4160 & 0.3320 & \textbf{0.4240}$\dagger$ & 0.4320 & 0.3440 & 0.3720 & 0.3880 \\\hline
TREC10 & 0.3520 & 0.4040 & 0.4720 & 0.3840 & 0.4520 & 0.4920 & 0.3800 & 0.4200 & 0.4880 \\\hline\hline
\multirow{2}{*}{\textbf{Pr@10}} & \multicolumn{3}{c|}{Dir} &    \multicolumn{3}{c|}{DirV} & \multicolumn{3}{c|}{JMV2} \\\cline{2-10}
 & SK & SV & LV & SK & SV & LV & SK & SV & LV \\\hline
TREC7  & 0.3980 & 0.4120 & 0.4420 & \textbf{0.4180}$\dagger$ & 0.4420 & \textbf{0.4720}$\dagger$ & 0.4100 & 0.4440 & \textbf{0.4800}$\dagger$ \\\hline
TREC8  & 0.4460 & 0.4120 & 0.4660 & \textbf{0.4740}$\dagger$ & 0.4380 & 0.4780 & \textbf{0.4700}$\dagger$ & 0.4400 & 0.4480 \\\hline
WT2G  & 0.4660 & 0.4220 & 0.4240 & 0.4840 & \textbf{0.4840}$\dagger$ & \textbf{0.4800}$\ddagger$ & 0.4920 & \textbf{0.4900}$\ddagger$ & \textbf{0.4820}$\ddagger$ \\\hline
TREC9  & 0.2560 & 0.2860 & 0.3160 & 0.2780 & \textbf{0.3260}$\ddagger$ & \textbf{0.3540}$\ddagger$ & 0.2780 & \textbf{0.3160}$\dagger$ & 0.3220 \\\hline
TREC10 & 0.3060 & 0.3500 & 0.4040 & 0.3300 & 0.3820 & 0.4340 & 0.3300 & 0.3700 & 0.4340 \\\hline
\end{tabular*}
\label{tbl_experimental_results}
\end{table}

\section{Conclusion}
This paper introduced a new issue for TF normalization by considering two different types of long-length documents - verbose documents and multi-topical documents. We proposed a novel TF normalization method which uses a partially-axiomatic approach. To this end, we formulated two desirable constraints, which the retrieval function should satisfy, and showed that previous language modeling approaches do not satisfy these constraints well. Then, we derived novel smoothing methods for language modeling approaches, without losing basic principles, and showed that the proposed methods satisfies these constraints more effectively.  Experimental results on five standard TREC collections show that the proposed methods are better than previous smoothing methods, especially for verbose type of query. JMV2 significantly improved JM for all type of queries, and DirV eliminated the limitation of Dir by providing the robustness of performances for verbose type of query, as well as improving precisions (Pr@5 or Pr@10) for keyword type of query. This is comparable to recent results using more complicated query-specific smoothing based on Poisson language model \cite{mei07}.

To handle long-length documents, passage-based retrieval could be applied \cite{kaszkiel01}. However, passage-based retrieval has a burden of decreasing efficiency, since it requires additional process such as indexing of position information, pre-segmenting individual passages, and more importantly the additional overhead at online retrieval time. Contrast to the complicated method such as the passage retrieval, this paper handles multi-topical documents in a simplified manner by investigating a more accurate TF normalization without additional cost of efficiency.



\mbox{}\newline
\textbf{Acknowledgement}. This work was supported by the Korea Science and Engineering Foundation (KOSEF) through the Advanced Information Technology Research Center (AITrc), also in part by the BK 21 Project and MIC \& IITA through IT Leading R\&D Support Project in 2007.

\bibliographystyle{splncs}
\bibliography{bibs_compact}

\end{document}